\documentstyle[12pt,epsf]{article}
\begin{document}
\newcommand{\bfa}{{\bf a}}
\newcommand{\bfb}{{\bf b}}
\newcommand{\bfc}{{\bf c}}
\newcommand{\apo}{a^{+1}}
\newcommand{\amo}{a^{-1}}
\newcommand{\az}{a^0}
\newcommand{\bpo}{b^{+1}}
\newcommand{\bmo}{b^{-1}}
\newcommand{\bz}{b^0}
\newcommand{\cpo}{c^{+1}}
\newcommand{\cmo}{c^{-1}}
\newcommand{\cz}{c^0}
\newcommand{\exk}{e^{ikr}}
\newcommand{\exa}{e^{-\alpha r}}
\newcommand{\bfr}{{\bf r}}
\newcommand{\bfq}{{\bf q}}
\newcommand{\bfk}{{\bf k}}
\newcommand{\bfR}{{\bf R}}
\newcommand{\bfp}{{\bf p}}
\newcommand{\hr}{\hat{\bf r}}
\newcommand{\ea}{{\it et al.}}
\newcommand {\eq}{\begin{equation}}
\newcommand {\qe}{\end{equation}}
\newcommand {\cen}[1]{\begin{center} #1 \end{center}}
\newcommand {\pr}{Phys. Rev.}
\newcommand {\pip}{$\pi^+$ }
\newcommand {\pim}{$\pi^-$ }
\newcommand {\h}{\frac{1}{2}}
\newcommand {\bfP}{{\bf P}}
\newcommand {\ppi}{{\bf p}_{\pi}}
\newcommand {\prl}{{\it Phys. Rev. Lett. }}
\newcommand {\pl}{{\it Phys. Lett. }}
\newcommand {\nucp}{Nucl. Phys. }
\newcommand {\prc}{Phys. Rev. C}
\newcommand {\de}{$^\circ$}
\baselineskip=14pt
\lineskip=15pt plus 0.2pt minus 0.2pt
\begin{center}

{\large \bf Pion Charge Exchange on Deuterium}\\

\vspace*{.15in}

                           J.-P. Dedonder\\
GMPIB\\
                Universit\'{e} Paris 7-Denis Diderot\\
                        2 Place Jussieu\\
                     F75251 Paris cedex 05, France\\

\vspace*{.1in}

                               and\\

\vspace*{.1in}
                          W. R. Gibbs\\
           Department of Physics, New Mexico State University\\
                       Las Cruces, NM 88003\\

\vspace{0.35in}
{\bf Abstract}
\end{center}

We investigate quantum corrections to a classical intranuclear cascade
simulation of pion single charge exchange on the deuteron. In order to
separate various effects the orders of scattering need to be
distinguished and, to that end, we develop signals for each order of
scattering corresponding to quasi-free conditions. Quantum corrections
are evaluated for double scattering and are found to be large. Global
agreement with the data is good.

\newpage

\section{Introduction}

The solution of the many-body Schr\"{o}dinger equation for scattering
problems is difficult indeed.  For this reason quantum mechanical reaction
calculations are often replaced with their classical analogs.
The replacement of the quantum problem by the classical one was 
first suggested by Serber \cite{serber}.  He observed that the early
data seemed to be consistent with a simple cascade of collisions
within a Fermi gas model of the nucleus. This idea was followed
by a long list of developing codes (see Refs.  \cite{goldberger}
through  \cite{gs}).

For heavy ion reactions the final state is very complicated and the
cascade calculation became one of the few tools available to predict
results. RQMD \cite{rqmd} and the Li\`ege code \cite{cugnon} provide
two standard calculational techniques for treating intermediate energy
heavy ion reactions. The code from Valencia \cite{valencia} is capable
of treating proton and pion projectiles.  New approaches include a
code developed by Li \ea\ \cite{li} using a set of coupled transport
equations and a cascade model developed with relativistic heavy-ion
collisions in mind, the ARC code \cite{kahana}.

Almost all of these models rely on the treatment of the scattering from
the point of view of classical probabilities with each scattering
being treated independently.  Of course we know that there are
phases arising from quantum mechanics which should enter into the
calculation of the total probability.  

If one wanted to perform a fully quantum mechanical cascade one
possibility might be to consider all possible results arising from the
initial conditions and then calculate the quantum mechanical probability
of each event. These probabilities would then be used as weights. This
would be an extremely inefficient procedure, however, since the dominant
fraction of the events, if chosen completely randomly, would occur with
very small probability and most of the calculational time would be wasted.  
A far more efficient procedure would be to choose the events according to
some approximate (well defined) probabilistic rule and then correct the
approximate rule by taking for the weight of the event the ratio of the
probabilities.  This is a standard technique in Monte Carlo procedures
known as ``importance sampling.'' The approximate event generator must
have the property that it can be sampled and that the probability for a
given event can be calculated. Of course, the full calculation will be
more efficient if the event generator gives results close to the
``correct'' answer. For this event generator we will use the classical
simulation mentioned above.  It is known to give good results for simple
reactions such as quasi-elastic scattering. Deviations from the model are
also seen and it is never sure if these discrepancies should be ascribed
to new physics or to the fact the model is purely classical.  Because of
its simplicity we have chosen the $\pi^++d\rightarrow pp\pi^0$ reaction to
investigate some quantum corrections.

To elucidate, to the extent possible, the role and magnitude of the 
quantum corrections we have been led to explore the role of various 
parameters (counter size, lower momentum cuts, absorption parameters etc.) 
either in relation to the data or the model.

\section{Data}

For the case of single charge exchange on the deuteron, there exist
fairly complete data \cite{tacik} on the cross section with a
coincidence between the two final protons such that the final state
was entirely determined. The coincidences were between pairs of
counters on opposite sides of the beam at 228 and 294 MeV so that the
reaction took place in a plane.  The counter positions are at 20\de,
45\de, 60\de\ and 125\de\ on each side of the beam. This would seem to
give 16 pairs but all coincidences were not possible. Removing the
(20\de,--20\de), (60\de,--125\de), (125\de,--60\de)  and
(125\de,--125\de) cases, there remain 12 angle pairs. By convention,
the momentum distribution displayed will be that of the counter
corresponding to the first element of the angle pair.

The coincidence requirement opens up new possibilities for analyzing the
data.  The quantum correction that we consider is the one due to quantum
vs. classical double scattering. The data of Tacik et al. \cite{tacik}
present the opportunity to explore its nature. One might think that the
double scattering contribution to single charge exchange on the deuteron
is small, and that is, in general, true.  However, the cases in which just
one charge exchange on the neutron takes place (without a scattering from
the other nucleon) the spectator proton has very low momentum in the final
state due to the low Fermi momentum of the deuteron.  In most experiments
(in particular the one we shall consider here \cite{tacik}) the two protons
are detected with a minimum momentum. Under these conditions the double
and higher scattering contributions are the most important.

This positive feature is balanced by the fact that the acceptance of the
experimental system must be understood.  Since the spectrum of one proton
was measured in coincidence with the second proton the threshold on the
second detector is important.

The data show a smooth variation as a function the counter pair position 
as well as a function of incident pion energy. However there are very 
strong, narrow peaks in the momentum spectra which are perhaps surprising. 
We shall argue that these peaks are normal kinematic features and can, to
some extent, serve as indicators of orders of scattering.
We need to characterize the data in terms of the multiple scattering
components since the relative weights of each order are different at 
228 and 294 MeV. 

In the treatment of this data we first run a cascade code to generate an
event file with charge exchange events. This file is then analysed with
appropriate threshold cuts, selection of the number of scatterings etc.
After some studies attempting to match the experimental thresholds we
decided to use a standard momentum threshold of 226 MeV/c for all of the
cases. The counters were taken to have an extension of 5\de\ in the
$\theta$ sense and the back-to-back condition was enforced by requiring
that $\cos\Delta\phi\le -0.99$, where $\Delta\phi$ is the difference in
azimuthal angles for the two protons.

\section{Classical code}

The present INC (Intra-Nuclear Cascade) code was originally developed
to treat moderate-energy antiproton annihilation in nuclei and has
been applied to that end several times
\cite{anti1,wrg,gs,fermilab,gkruk}. However, the annihilation of an
antiproton leads to pions (or at least it is so treated by the model)
and so the history of pions in the energy range below, and of the
order of, 1 GeV is essential to the calculation of energy deposition.  
Of particular importance for the thermalization of the nuclear matter
are pion absorption and production.  For this reason the code needed
to be checked against reactions initiated by pion beams \cite{light}.  
Calculations have been done to compare with data on antiproton
annihilation of 5-10 GeV antiprotons on several nuclei \cite{rice}.  
Considerable success has been obtained in predicting the rapidity
distributions of strange particles produced in antiproton reactions
\cite{gkruk}. The question of pion absorption and comparison with data
has been addressed \cite{lanl} and the code has been used for the
comparison with inclusive data \cite{zumbro}.  It was quite successful
in describing the overall spectrum although there is a problem with
the number of final pions in the region of the delta resonance.

A description of the basic features of the pion version of the code can be 
found in Ref.  \cite{alqadi} where pion double charge exchange on
$^4$He was treated with a quantum correction for final state interaction
for the (unobserved) nucleon pair on which the two charge exchanges took
place. This was an important correction since, for high
energy final pions, these two nucleons often have low relative energies.
In the present case the two final nucleons cannot have low relative
energies (because the counter pairs require a substantial separation
in angle) so this final-state-interaction correction was not included.

As can be seen in the above mentioned descriptions, the absorption is
controlled by a parameter which expresses the probability that an
absorption takes place when it is permitted by conservation laws. In
the present work we varied this parameter to fit the experimental
absorption cross sections  \cite{ritchie}.  Even large variations of this
cross sections (a factor of two) led to variations in the magnitude of the
charge exchange cross sections of at most of the order of 20-30\% with no
major change in the shape of the spectra.

The initial bombarding pion is started with the appropriate initial
momentum toward a circle (in this case with a radius of 6.4 fm) which is
large enough to contain the projected density of the target nucleus.  A
fraction of the beam particles (usually about one half) pass without
interacting and cross sections are computed as the fraction of the
reactions of interest which occur multiplied by the area of this disk.

The first problem to be approached in attempting to make a classical
solution to the many-body scattering system act as a simulation of the
quantum system is the realization of the initial density of particles
in the bound target.  While one can choose the coordinate positions
appropriately, the distribution of momenta of the particles (we
will be mainly interested in magnitudes since the s-wave nature of the
present problem means that the directions of the momenta will be
isotropic) also needs to be taken to match the quantum case.  The
technique for the construction of a nucleus with A nucleons is given
in Ref. \cite{alqadi}.

For the case of the deuteron we can take the spatial distribution
directly from a probability density. Since there are two particles
with the center of mass located at the origin and the location of
the particles is isotropic, a direct sampling method can be used.
For the radial density we have taken that of the solution with
a one-pion-exchange potential \cite{rosa,fg,ballot,book}.

Since we intend to have the two nucleons propagate under the action
of a potential, once the position of an initial nucleon is 
established the kinetic energy is fixed by the relation
\eq
T(r)+V(r)=E
\qe
where $T$ is the kinetic energy of the particle and $V$ is the
potential chosen.  Once a value of $r$ is chosen then a value of the
kinetic energy (and hence of momentum) is fixed.  Since (as we will
see shortly) the coincidence requirement leads to delta functions for
the momentum distributions for double scattering in the absence of
Fermi motion in the deuteron it is important that the distribution of
this quantity be realistic.

If the potential in this equation is chosen to be the same as that
used in the quantum mechanical problem to provide a solution giving
the density, in general the distribution of the momenta obtained from
the classical procedure just outlined will give a (very) different
distribution of momenta from the one obtained from the square of the
momentum space wave function derived from the solution of the quantum
problem. In particular, the quantum solution gives a distribution of
momenta which has support to infinity whereas the classical solution,
because of the fact that a typical potential used in the solution of
the nucleon-nucleon problem has a maximum depth, has a cut-off at a
finite value.  This cut-off comes at a point well within the range of
interest of momenta so that the resulting momentum distribution is far
from realistic.

This problem can be solved (at an expense as we shall see) by choosing the
potential such that the momentum distribution is correct if the radial
density is the one desired.  In the present case, we have taken the
binding energy of the deuteron to be zero so that the kinetic energy is
equal to the negative of the potential.

\begin{center}\begin{figure}[p]
\epsfysize=185mm
\epsffile{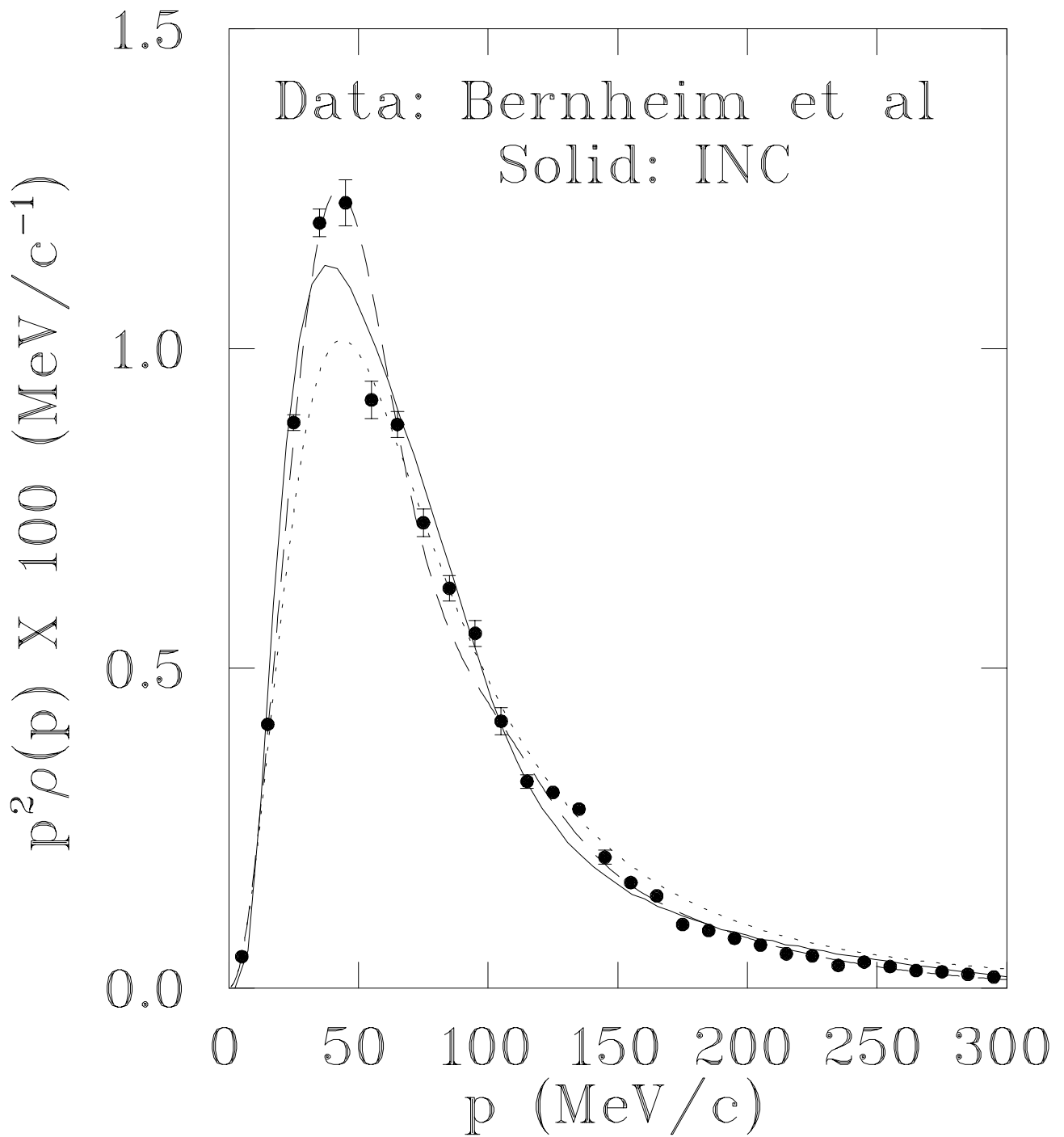}
\caption{Comparison of the measurements of Fermi momenta by Bernheim
et al.\protect \cite{bern} with a fit to the data (dashed line) and the 
result of the INC (solid line) using the potential derived in the text and 
shown
in Fig. \protect\ref{vbern}. Also shown is the square of the momentum
wave function of the one-pion-exchange deuteron (dotted).}
\label{bern}\end{figure}\end{center}

To find the potential which will make these two distributions 
compatible in the classical sense, we first transform the 
momentum distribution to a distribution of kinetic energies. The 
momentum distribution used in this case is taken from a fit to the 
data of Bernheim et al. \cite{bern} where the data and the fit are
shown in Fig. \ref{bern}.

Given this distribution, the condition that the kinetic energy
distribution, $g(T)$, be obtained from a given radial distribution,
$\rho(r)$, is

\eq
\rho(r)dr=g(T)dT
\qe
for which the integrated form (taking account of the proper limits
to give the boundary conditions) is
\eq
F(r)\equiv \int_r^{\infty}\rho(r)dr=\int_0^Tg(T)dT\equiv G(T).
\qe
To obtain the desired solution the functions $F(r)$ and $G(T)$ are
tabulated numerically and then the numerical inversion
\eq
T(r)=G^{-1}[F(r)]
\qe
is made. The potential is then identified with the negative of the kinetic
energy.  The numerical inversion procedure introduces some error but is
stable except for very small values of r where the numerical procedure
limits to a constant potential whereas the true result goes to infinity. A
fit is then made to the potential which follows the potential in the
region where it is well determined. In general the procedure works well
and the resultant momentum distribution from the simulation is shown in
Fig. \ref{bern} compared with the input distribution. The agreement is
good but not perfect.

\begin{center}\begin{figure}[p]
\epsfysize=185mm
\epsffile{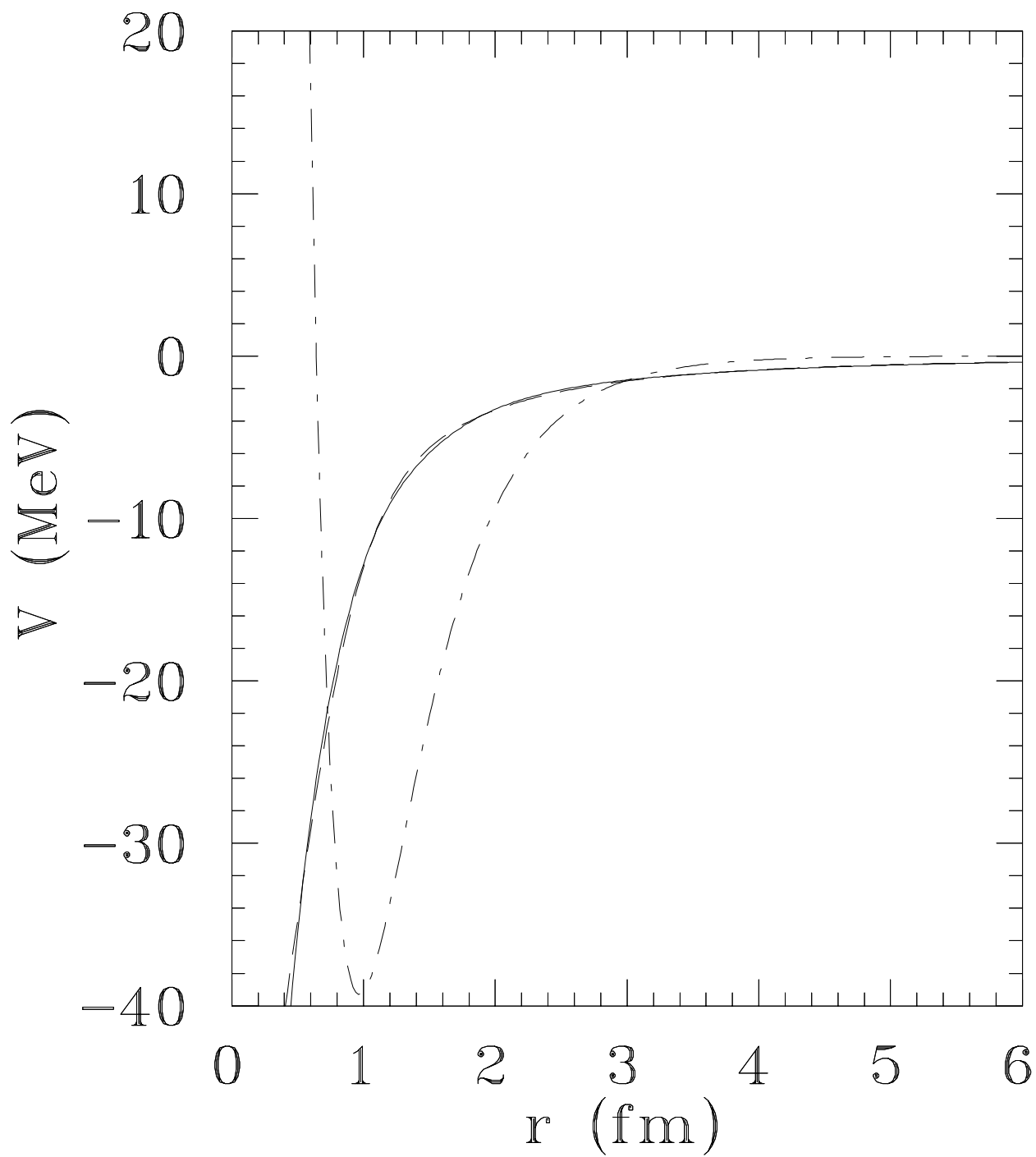}
\caption{The potential obtained as described in the text. The solid
line is the potential directly from the procedure and the dashed line
shows the fit used in the cascade code.  The dash-dot line shows the
Malfliet-Tjon \cite{mt} potential as modified \cite{book}.}
\label{vbern}\end{figure}\end{center}

The resulting potential is shown in Fig. \ref{vbern}. It is seen
that it has only a cursory resemblance to a semi-realistic NN
potential at large r and is completely different at small r where it
lacks the repulsive hard core.  This potential is the price paid for
being able to have correct spatial and momentum distributions with
conservation of energy. For double scattering or higher this potential
is not very important since the nucleons have high energies and are
little affected by the final state interaction between the two
nucleons.

For single scattering, however, where the Fermi momentum (after final
state interaction) must be detected in one of the pair of counters,
the error can be substantial.  Since only the tail of the Fermi
momentum distribution has large enough values of momentum to be
observed in the detectors, the exact values of the momentum in the
tail is crucial. For a number of cases the single scattering plays a
small role while for others it contributes in certain parts of the
spectrum in ways that might not be imagined without some reflection.  
In the case where the original proton (not struck) is in the backward
direction (for one of the counters at 125\de\ in the present
experiment) the force due to the final state interaction will act to
slow it down.  If both nucleons are going forward in the final state
the force acts so as to increase the momentum of the spectator proton.  
Since the potential obtained with this procedure is too strong for
high momenta (compared with a realistic one) these effects may be over
estimated.

In the case where the momentum is determined by a potential, as in the
present model, we can get some feeling for this correction
(independent of the form of the potential) by considering a simplified
case. We assume that particle two (the spectator particle)  has a
given Fermi momentum $\bfp_2$ with the angle being given by the
coincidence counter.  Then particle one must have an initial momentum
equal and opposite.

With the usual center of mass expressions
\eq
\bfp=\frac{\bfp_1-\bfp_2}{2};\ \ \bfP=\bfp_1+\bfp_2;\ \ 
\bfp_1=\frac{\bfP}{2}+\bfp;\ \ \bfp_2=\frac{\bfP}{2}-\bfp
\qe
where, in fact, $\bfP=0$ in this case, we can write the sum of the 
kinetic and potential energy in the initial state as
\eq
\frac{\bfp^2}{m}+V=0,
\label{initial state}
\qe
where we assume zero binding. After the scattering with momentum
transfer, $\bfq$,\ we have
\eq
\bfp'_2=\bfp_2;\ \ \bfp'_1=\bfp_1+\bfq;\ \ \bfp'=\bfp+\frac{\bfq}{2}.
\qe
In the final state the relative momentum at infinity will be given by
\eq
\frac{{\bfp'}^2_{\infty}}{m}
=\frac{{\bfp'}^2}{m}+V=
\frac{\bfp'^2}{m}-\frac{\bfp^2}{m}\ \ {\rm or}\ \ 
{\bfp'}_{\infty}^2=\bfp'^2-\bfp^2=\frac{\bfq^2}{4}+\bfq\cdot\bfp.
\qe
Assuming that the direction in the center of mass does not change
as the particles propagate to infinity (as would be the case when
they are back to back)
\eq
\bfp'_{\infty}=\sqrt{\frac{\bfq^2}{4}+\bfq\cdot\bfp}\frac{\bfp'}
{|\bfp'|}=
\frac{\sqrt{\frac{\bfq^2}{4}+\bfq\cdot\bfp}}
{\sqrt{\frac{\bfq^2}{4}+\bfq\cdot\bfp+\bfp^2}}(\bfp+\frac{\bfq}{2}),
\qe
the final momentum of the spectator particle in the laboratory
will be
\eq
\bfp'_{2\infty}=\frac{\bfq}{2}+\frac{\sqrt{\frac{\bfq^2}{4}+\bfq\cdot\bfp}} 
{\sqrt{\frac{\bfq^2}{4}+\bfq\cdot\bfp+\bfp^2}}(\bfp_2-\frac{\bfq}{2}).
\qe
Thus, as $|\bfq|\rightarrow\infty$ the momentum of the spectator in the 
final state becomes equal to the initial Fermi momentum. However, 
the present case treats only moderate values of momentum transfer 
so a substantial correction is to be expected.

\begin{figure}[p]
\begin{center}
\epsfysize=185mm
\epsffile{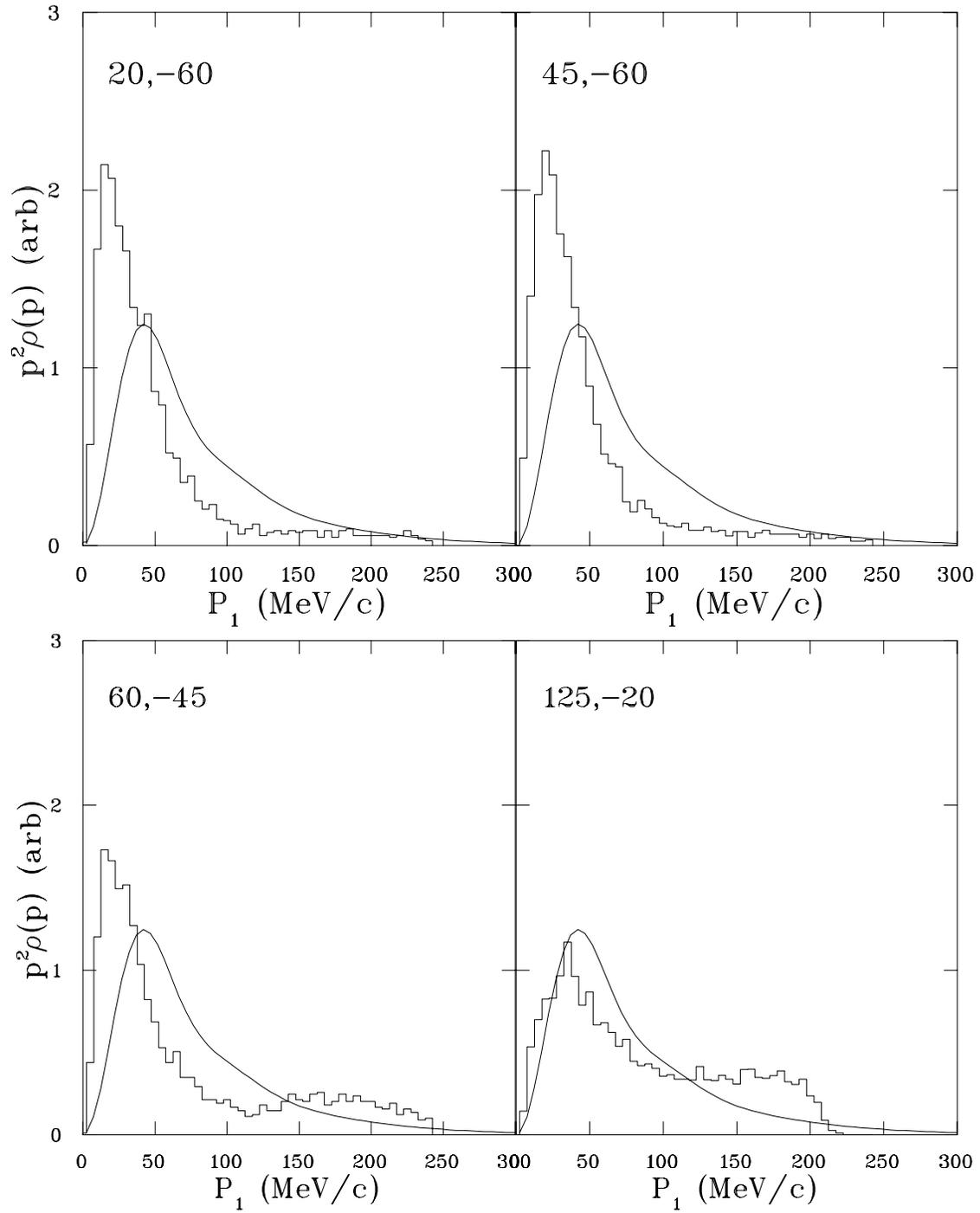}
\caption{Comparison of the observed final momentum of the
unstruck particle in single scattering events with the initial
Fermi momentum (its initial momentum before interaction with the
potential.}
\label{specmoma}
\end{center}
\end{figure}

Since we know the initial distribution and we can select the events with
single scattering in the calculation and accumulate the distributions of
the final proton momenta, the effect of this potential in the final state
can be observed.  Figure \ref{specmoma} shows such a comparison for four
angle pairs. The curves have been normalized to the same integral values.  
It is seen that the momenta for the case of the forward angle counters
have been shifted to lower values as expected from the above arguments.  
In other cases the distribution is very similar to the starting
distribution or increased at high momenta. The largest angle counters show
a double peaked structure.

\section{Single, double and triple scattering}

While quasi-free single scattering peak has been known for a long
time, it is interesting that in a coincidence experiment, one can
expect peaks from quasi-free double and triple scattering. Since peaks
in spectra are sometimes interpreted as particle masses, one should be
aware of the possible presence of such peaks to avoid
misinterpretations which could arise. Scattering is carried only to
4th order, i.e. after the pion has scattered four times it is not
allowed to interact. Thus, what we call quadruple scattering really
represents all of the rest of the scatterings which would have
occurred as well. We are not able to give an analytical discussion of
this higher order but we will treat the first three orders.

\subsection{\bf Single scattering}

The quasi free scattering peak is well known in measurements in which a
single particle is observed and, indeed, appears prominently in many
cases. It corresponds to a free nucleon at rest being struck. Since the
Fermi momentum distribution typically peaks at zero (or low) momentum, a
peak in the final momentum distribution is observed at the value of 
momentum appropriate for free scattering. In addition there is 
a distribution of counts on either side with the extent of the wings
depending on the Fermi momenta.

In the present coincidence experiment, where for the case of single
scattering, a substantial Fermi momentum is needed for the
observation of the spectator proton, the maximum of the quasi-free
peak is explicitly excluded. The most one might expect to see is one
or both of the wings of the distribution. This effect can lead to
rather unexpected contributions to the spectrum.

\begin{figure}[p]
\begin{center}
\epsfysize=185mm
\epsffile{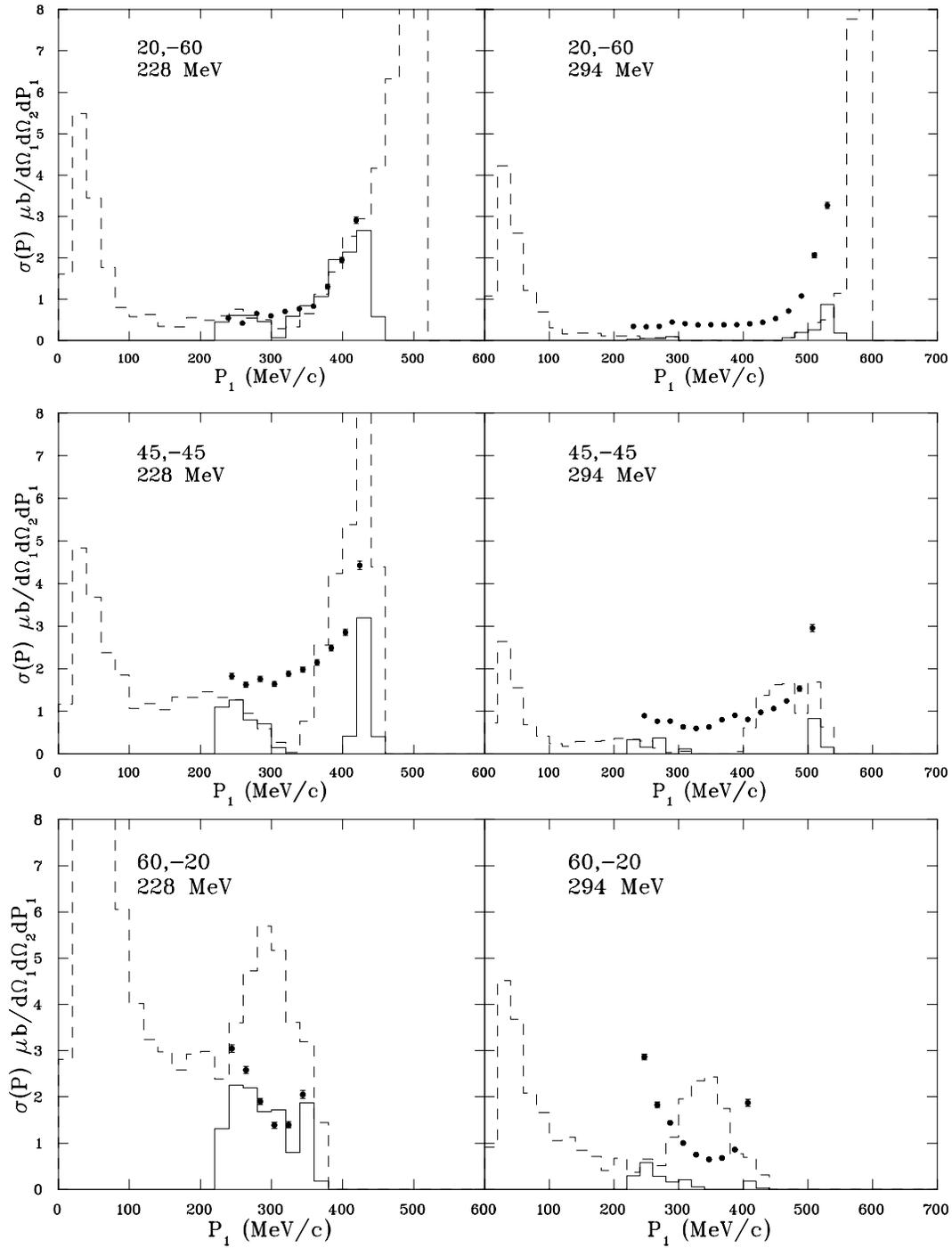}
\caption{Single scattering with and without cuts at 228 and 294 MeV.
The data points shown in this figure and all of the figures to follow
is from \protect{ \cite{tacik}}.}
\label{sbernfca6}
\end{center}
\end{figure}

Figure \ref{sbernfca6} shows results for single scattering.  The
dotted curve shows the distribution without any thresholds for the
counters, and the quasi-free peak is clearly seen.  The solid curve
displays the result with the threshold cuts in place and one sees that
most of the single scattering is eliminated by the cuts. In some cases a
remnant of the single scattering is left. Interesting are the cases of the
angle pairs (45\de,--45\de) and (60\de,--20\de) where the quasi-free peak
is in the center of the spectrum and only the tails of the distribution
remain after the cut resulting in peaks at high and low momenta, with
precisely the opposite shape to the original spectrum before the cuts.  
Clearly, it is difficult to be sure of the strength of these peaks since
they depend on the values of the cuts and, especially, on the final state
interaction potential. In these two cases, since the momentum distribution
has been modified only slightly by the final state interaction one may
expect that the predictions are at least qualitatively correct. 

The remnants after cuts shown in Fig.\ \ref{sbernfca6} for 294 MeV are
among the largest for that energy. The remaining single scattering cross
sections at 228 MeV are large, not only in the case of the counter pairs
shown but in the pairs (20\de,--45\de) and (45\de,--20\de) and to a lesser
extent for the pairs (45\de,--60\de) and (60\de,--45\de). For the
(20\de,--60\de) angle pair the final momentum for the quasi-free peak is
clearly visible without cuts but mostly eliminated with them.  The effect
of the cuts is rather different at 228 MeV and 294 MeV.

\subsection{Double scattering}

We now discuss the existence of quasi-free double scattering peaks
where each of the two particles will receive a substantial momentum
from the scattering process. For this reason in this study we limit
ourselves to the case of zero Fermi momentum.  By specifying the angle
of the outgoing (first)  nucleon, with the incident pion momentum
known, the kinematics of the reaction are expressed by
\eq
\bfk_{\pi}=\bfk_1+\bfk 
\qe 
where $\bfk_1$ is the final energy of the first
struck nucleon and $\bfk$ is the pion momentum after the first scattering.
Equating the total laboratory energy before and after scattering we have:
\eq 
E=\omega+M= \sqrt{\mu^2+(\bfk_{\pi}-\bfk_1)^2}+\sqrt{M^2+k_1^2}
=\sqrt{\mu^2+k_{\pi}^2+k_1^2-2k_{\pi}k_1x}+\sqrt{M^2+k_1^2} 
\qe 
where $x$ is the cosine of the angle between the incident pion 
direction and $\bfk_1$. Solving this equation for $|\bfk_1|$ we have 

\eq 
|\bfk_1|=k_1=\frac{2M k_{\pi} x}{E-k_{\pi}^2x^2/E}.
\qe 

Since the final pion momentum from the first scattering is known, it
can be used as input for the second scattering and, with the direction
of the final nucleon fixed by the experimental conditions, all angles
and energies are again known. We can apply the same formula to find
\eq
k_2=\frac{2M k y}{E'-k^2y^2/E'}
\qe
where $E'=\sqrt{k^2+\mu^2}+M$ and $y$ is cosine of the angle between 
$\bfk$ and $\bfk_2$.

Thus, for a given angle pair there are two momenta (each in a different
counter) where one might expect to observe a peak.  Since the first
scattering must lead to the recoil of the nucleon in the forward
direction, when one counter at 125\de\ is involved there is only one
value possible corresponding to the scattering to the forward counter
first. Tables \ref{dtable294} and \ref{dtable228} give the peak position
expected at 294 and 228 MeV respectively.

\begin{table}[htb]
$$\begin{tabular}{|c|c|c|c|c|}
\hline
{\rm Angle\ Pair}&20,\ 125&20,\ 60\ &20,\ 45\ &\\
\hline
{\rm Peak\ Position(s) (MeV/c)}&573\ \ \ \ \ &573\ \ 507&573\ \ 316&\\
\hline
\hline
{\rm Angle\ Pair}&45,\ 125&45,\ 60\ &45,\ 45\ &45,\ 20\ \\
\hline
{\rm Peak\ Position(s) (MeV/c)}&415\ \ \ \ \ &415\ 557&415\ 447&415\ 75\ 
\\
\hline
\hline
{\rm Angle\ Pair}&&60,\ 60\ &60,\ 45\ &60,\ 20\ \\
\hline
{\rm Peak\ Positions (MeV/c)}&&287\ 529&287\ 486&287\ 165 \\
\hline
\hline 
{\rm Angle\ Pair}&&&125,\ 45&125,\ 20\\
\hline
{\rm Peak\ Position (MeV/c)}&&&\ \ \ \ \ \ 259&\ \ \ \ \ \ 376\\
\hline
\end{tabular}
$$
\caption{Expected peaks from quasi-free double scattering for a pion
incident energy of 294 MeV. The numbers are for the position of peaks
expected in the first counter of the pair. The first number corresponds
to the case where the first struck nucleon was detected in this counter
and the second number corresponds to the case where the second
scattered nucleon was detected in the first member of the counter pair.  
In the pairs in which one of the counters is at 125\de, only one value is
possible since the first scattering cannot lead to a particle recoiling
at greater than 90\de (without Fermi motion). }
\label{dtable294}\end{table}

\begin{table}[hbt]
$$\begin{tabular}{|c|c|c|c|c|}
\hline
{\rm Angle\ Pair}&20,\ 125&20,\ 60\ &20,\ 45\ &\\
\hline
{\rm Peak\ Position(s) (MeV/c)}&488 \ \ \ \ \ &488\ \ 427&488\ \ 260&\\
\hline
\hline
{\rm Angle\ Pair}&45,\ 125&45,\ 60\ &45,\ 45\ &45,\ 20\ \\
\hline
{\rm Peak\ Position(s) (MeV/c)}&357\ \ \ \ \ &357\ 476&357\ 378&357\ \ 
54\\
\hline
\hline
{\rm Angle\ Pair}&&60,\ 60\ &60,\ 45\ &60,\ 20\ \\
\hline
{\rm Peak\ Positions (MeV/c)}&&249\ 458&249\ 416&249\ 135 \\
\hline
\hline 
{\rm Angle\ Pair}&&&125,\ 45&125,\ 20\\
\hline
{\rm Peak\ Position (MeV/c)}&&&\ \ \ \ \ \ 241&\ \ \ \ \ \ 334\\
\hline
\end{tabular}
$$
\caption{Expected peaks from quasi-free double scattering for a
pion incident energy of 228 MeV. See Table \protect\ref{dtable294} for 
an explanation of the entries.}\label{dtable228}
\end{table}

\begin{center}
\begin{figure}[p]
\epsfysize=185mm
\epsffile{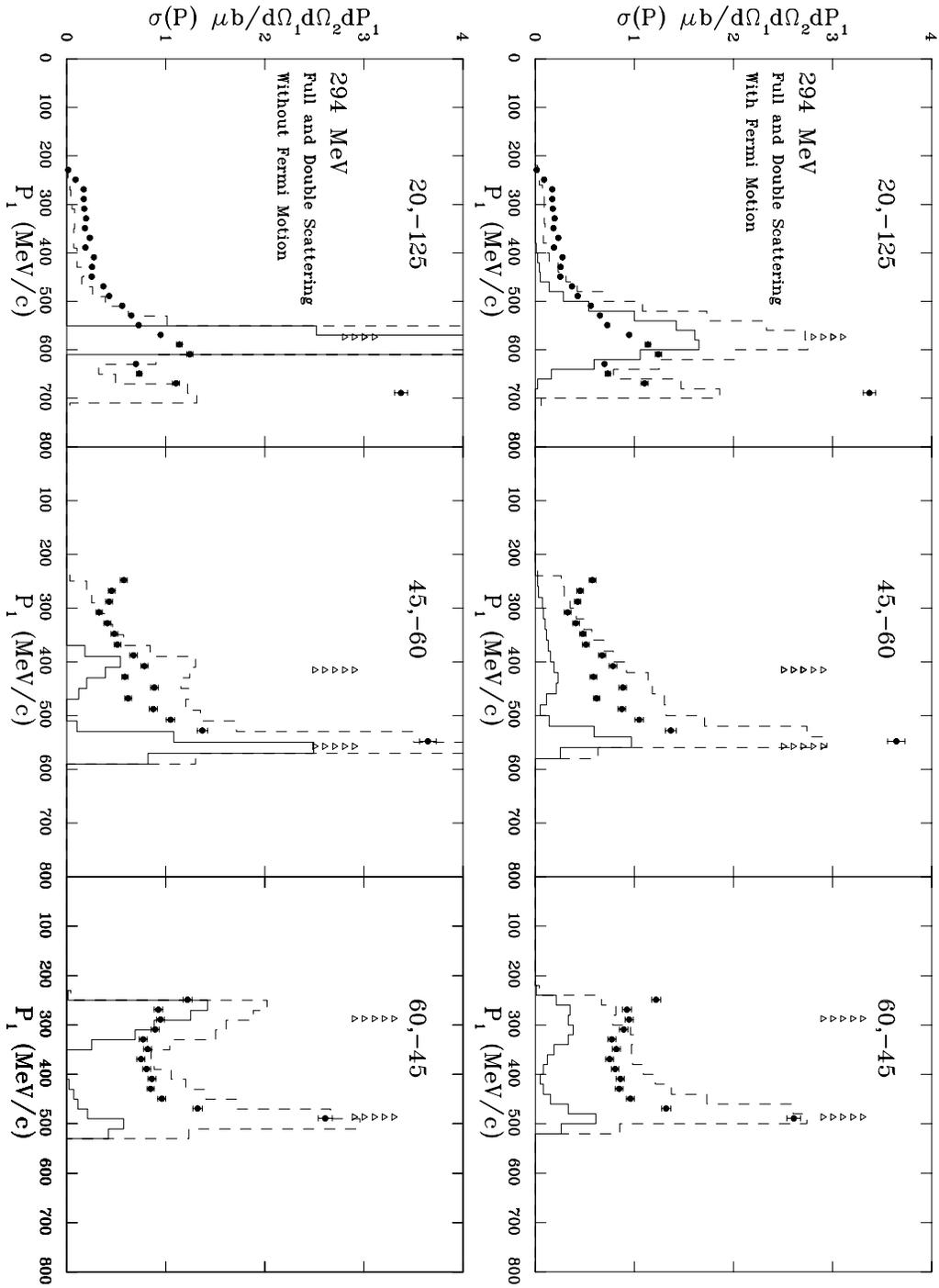}
\caption{Total and double scattering with and without Fermi motion at 294 
MeV. The solid line represents the contribution of double scattering
and the dashed line gives the total. Both calculations were made without 
absorption.}
\label{totbernfnf}
\end{figure}
\end{center}

Figure \ref{totbernfnf} compares calculations with and without
Fermi momentum for double and total scattering. They are made including
the quantum correction to be discussed in the next section. It is
seen that peaks do indeed come where predicted by the above 
considerations (shown as triangles in the figure). Fermi motion
and higher order scatterings tend to blur and hide them but they are
often visible in the final result.

\subsection{Triple scattering}

In this case (perhaps remarkably) one also has regions of strength in
the quasi-free process. The reason for the existence of
structure is a Jacobian peak introduced by a transformation discussed
in the following. If we assume that the entire triple scattering
remains in a plane, then for a given value of the recoil angle for the
initial scattering, $\theta^{i}_1$, for a fixed $\theta_2$ all
kinematics are defined. The probability of such an event will be given
in terms of a product of the three scattering cross sections involved.  
Performing the transformation from the distribution in $z=\cos
\theta^{i}_1$ to the distribution in final momenta, $k_1(z)$ (the
momentum of the first nucleon {\it after} the {\it second}
scattering), the momentum distribution is given by

\eq
\frac{dP}{dk_1}=\frac{dP}{dz}/|\frac{dk_1}{dz}|
\qe
where the quantity $dP/dk_1$ is the probability of the triple
scattering taking place for a given $z$. $dk_1/dz$ typically has a
zero in the range of interest. This zero occurs at the maximum energy
possible for triple scattering which, in fact, coincides with the
maximum energy possible for the reaction (regardless of the number of
scatterings). Triple scattering is the first order in which this
maximum momentum can be reached. This peak will have the same form for
any value of Fermi momentum and hence is not broadened by the motion
of the nucleon. Since the measurement is a coincidence cross section,
one expects a companion peak in the second counter at the energy of
the second scattering which corresponds to the Jacobian peak. While
the Jacobian peak is clearly seen in the experimental results the
companion peak is usually much broader and generally not visible. It
is worthwhile to note that the counter size can influence what is seen
since there is a true singularity in these peaks. Table \ref{ttable}
gives the positions of these Jacobian peaks and the companion peak.

\begin{table}[thb]
$$
\begin{tabular}{|c|c|c|c|c|}
\hline
&\multicolumn{2}{|c|}{294\ {\rm MeV}}&\multicolumn{2}{|c|}{228\ {\rm MeV}}\\
\hline
{\rm Angle\ Pair}&$k_1$ {\rm (MeV/c)}&$k_2\ {\rm (MeV/c)}$& 
$k_1$ {\rm (MeV/c)}&$k_2\ {\rm (MeV/c)}$ \\
\hline
20,\ -125&696&606&598&508\\
\hline
45,\ -125&493&419&440&379\\
\hline
20,\ -60&585&237&498&199\\
\hline
45,\ -60&557&336&477&285\\
\hline
60,\ -60&551&402&475&343\\
\hline
20,\ -45&576&144&490&117\\
\hline
45,\ -45&517&227&441&191\\
\hline
60,\ -45&499&286&428&239\\
\hline
125,\ -45&259&191&244&189\\
\hline
45,\ -20&462&38&394&28\\
\hline
60,\ -20&419&83&358&66\\
\hline
125,\ -20&377&287&335&252\\
\hline
\end{tabular}$$
\caption{Expected peaks in triple scattering at 294 and 228 MeV based on the 
position of the Jacobian singularity. Both entries at a given energy 
correspond to the momentum in the first member of the angle pair. 
}\label{ttable}
\end{table}

\begin{figure}[p]
\begin{center}
\epsfysize=185mm
\epsffile{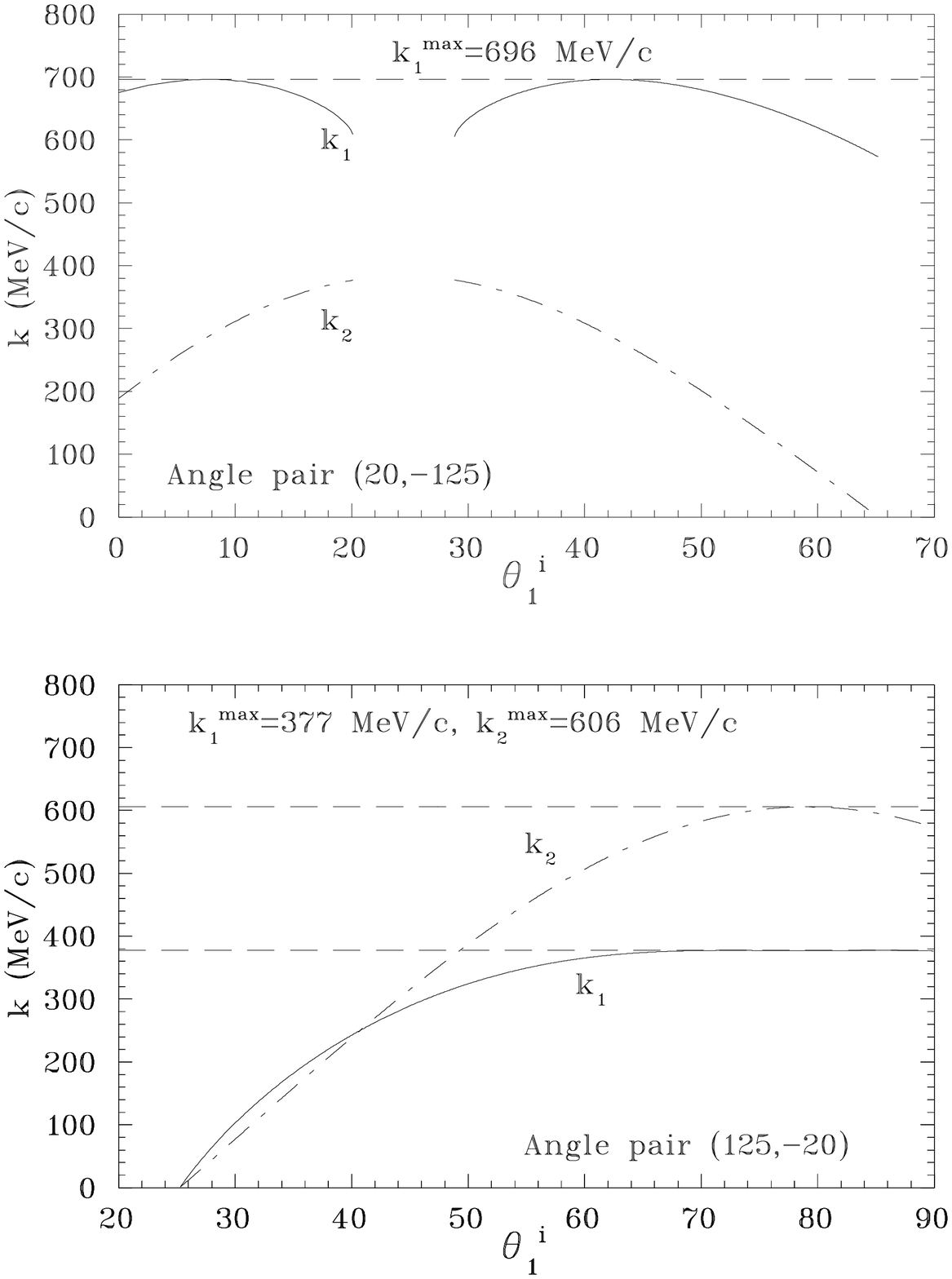}
\caption{Momenta for triple scattering at 294 MeV. The momentum $k_1$ 
is the one measured in the 20\de\ counter and $k_2$ that measured in 
the 125\de\ counter.}
\label{triple}
\end{center}
\end{figure}

Figure \ref{triple} gives the final momenta of the two nucleons as a
function of the (assumed in plane) scattering angle in the first
scattering for the angle pairs (20\de,--125\de) and (125\de,--20\de).
It is seen that the momentum of the particle in the second counter
also has a maximum (and hence also a Jacobian peak) for the case
(125\de,--20\de). For the conjugate pair, (20\de,--125\de) the case is
not realized for the peak of the second momentum.  However, this
second Jacobian peak in the 20\de\ counter means that there should be
two sharp peaks with no Fermi momentum. When Fermi momentum is
included in the problem the peak in the interior of the distribution
will be broadened but that at the maximum of momentum will not.

\begin{center}
\begin{figure}[p]
\epsfysize=185mm
\epsffile{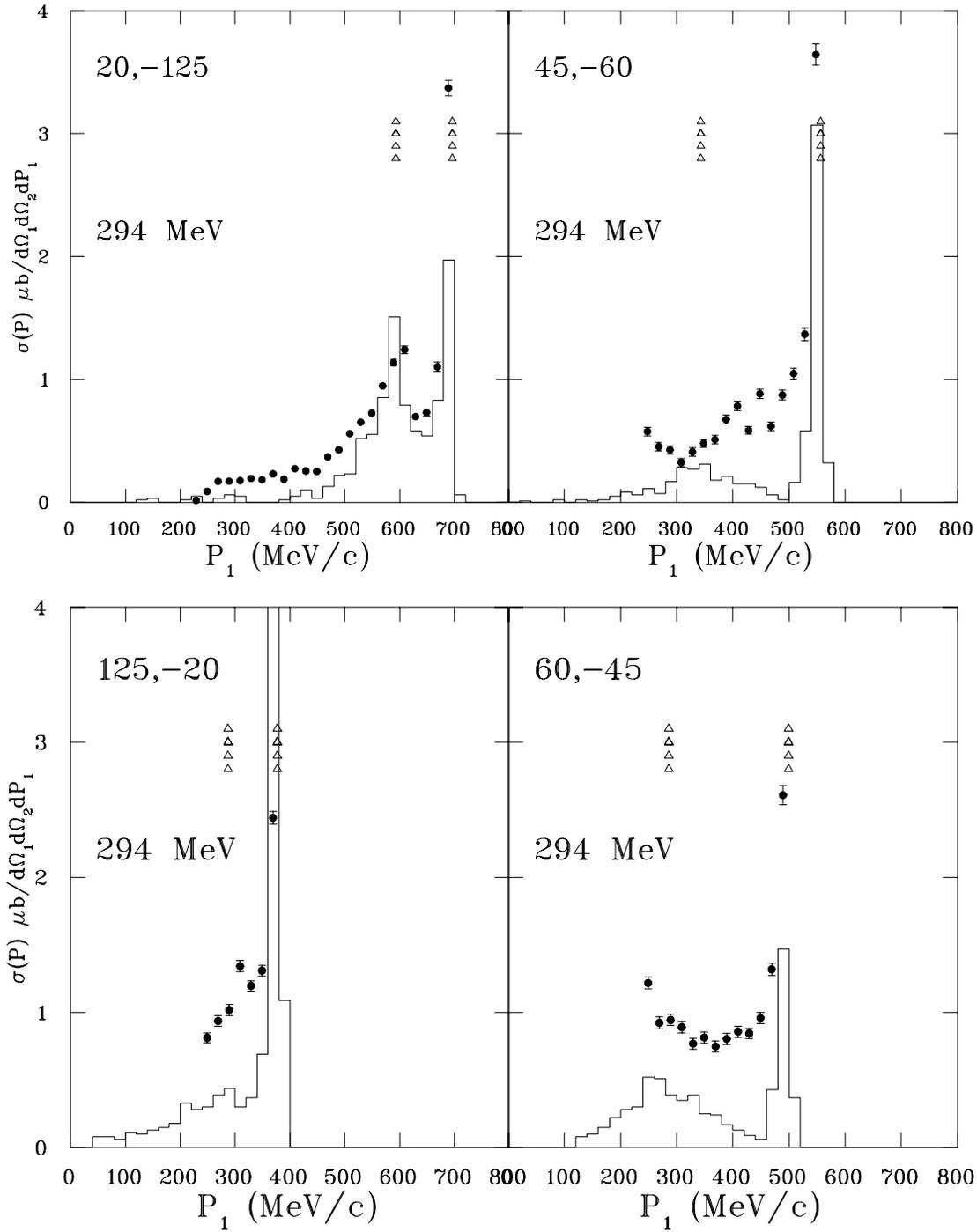}
\caption{Triple scattering without cuts, with absorption and with
a small Fermi momentum at 294 MeV to show the kinematic effects of the
Jacobian peaks.}
\label{tberna2}
\end{figure}
\end{center}

Figure \ref{tberna2} shows results of the INC calculation with a very
small Fermi momentum. It is seen that the peaks match the predictions
(marked with the triangles). While the companion peak to the Jacobian is
normally broad (see pair 45\de,--60\de), we see that, indeed, the 
angle pair (20\de,--125\de) is an exception with the second peak being 
also narrow. There is some broadening of the peaks due to the finite 
size taken for the counters in the analysis of the events coming from 
the INC.
 
\section{Quantum corrections}

In this section we discuss the quantum corrections that we apply
for the double scattering only. We will treat spin, space and isospin
in turn starting with the general form of the operator in spin space.

\subsection{General form}

In order to calculate the ratio of the quantum double scattering
cross section to the classical version we must evaluate the
double scattering amplitude.  Following the technique of Ref.\  
\cite{gek}
we can express the amplitude for a fixed position of the two nucleons
as an operator on a single function $g(r)$.

\eq 
e^{i(\bfk_1\cdot\bfr_1+\bfk_2\cdot\bfr_2-\bfk_{\pi}\cdot\bfr_1+
\bfk'_{\pi}\cdot\bfr_2)}
A_{ds}=(A_1+B_1\bfq\cdot\bfk+C_1\sigma_1\cdot\bfq\times\bfk)
(A_2+B_2\bfk'\cdot\bfq+C_2\sigma_2\cdot\bfk'\times\bfq )g(r) 
\label{adsdef} 
\qe
where $\bfq$ is to be interpreted as $-i\nabla$.  Here the constants
A, B and C are determined from pion-nucleon phase shifts and
correspond to the two processes possible: $\pi^+$ elastic scattering
on the proton followed by charge exchange on the neutron or charge
exchange on the neutron followed by $\pi^0$ elastic scattering on the
proton.  The phase factor on the left could be ignored in many cases but
we need to keep it here since we wish to consider the coherence of this
scattering (nucleon 1 followed by nucleon 2) with the reverse order.

The function $g(r)$ is given by
\eq
g(r)=\frac{e^{i\kappa r}-e^{-\alpha r}}{r}-\frac{(\kappa^2+\alpha^2)}
{2 \alpha}e^{-\alpha r}
\qe
where we have taken 
\eq  v(q)=\frac{\alpha^2+\kappa^2}{\alpha^2+q^2}, \qe
and $\kappa$ is the momentum of the intermediate propagating pion
and $\alpha$ is the range of the form factor (taken as 4 fm$^{-1}$ here).

We consider the transformations of the radius vectors according
to:

$$\bfr_1=\bfR +\frac{\bfr}{2}, \ \ \ \bfr_2=\bfR -\frac{\bfr}{2}.$$

We see that we have terms with no, one and two 
derivatives. Since $\nabla g(r)=\hr g'(r)=\bfr g'(r)/r$
we can make a simple replacement in the terms with one derivative.
For the terms with two derivatives a second term appears which 
corresponds to the operation of the derivative on the factor $\bfr$. 
Thus we can expand Eq. \ref{adsdef} as:
\eq e^{i(\bfk_1\cdot\bfr_1+\bfk_2\cdot\bfr_2-\bfk_{\pi}\cdot\bfr_1+
\bfk'_{\pi}\cdot\bfr_2)}
A_{ds}= A_1A_2\ g(r) \label{expand} \qe
$$ +(A_1B_2\ \bfk'\cdot\bfq +A_2B_1\ \bfq\cdot\bfk )\ g(r) $$
$$ +(B_1B_2\ \bfq\cdot\bfk\ \bfk'\cdot\bfq)\ g(r) $$
$$ +(A_1C_2\ \sigma_2\cdot\bfk'\times\bfq+A_2C_1\ \sigma_1\cdot\bfq
\times\bfk )\ g(r) $$
$$ +(B_1C_2\ \bfq\cdot\bfk\ \sigma_2\cdot\bfk'\times\bfq+
B_2C_1\ \bfk'\cdot\bfq\ \sigma_1\cdot\bfq\times\bfk) g(r) $$
$$ +C_1C_2\ \sigma_1\cdot\bfq\times\bfk\  
\sigma_2\cdot\bfk'\times\bfq)\ g(r), $$
where we have separated the terms according to the number of
derivatives and the number of occurrences of the spin operators.
Performing the operations we can write (still as an operator in
spin space):

\eq e^{i(\bfk_1\cdot\bfr_1+\bfk_2\cdot\bfr_2-\bfk_{\pi}\cdot\bfr_1+
\bfk'_{\pi}\cdot\bfr_2)}
A_{ds}= A_1A_2\ g(r) \label{expan} \qe
$$ -i(A_1B_2\ \bfk'\cdot\hr +A_2B_1\ \hr\cdot\bfk )\ g'(r) $$
$$ -B_1B_2\ \hr\cdot\bfk\ \bfk'\cdot\hr\ g^-(r)
 -B_1B_2\ \bfk\cdot\bfk'\ \frac{g'(r)}{r} $$
$$ -i(A_1C_2\ \sigma_2\cdot\bfk'\times\hr
+A_2C_1\ \sigma_1\cdot\hr\times\bfk )\ g'(r) $$
$$ -(B_1C_2\ \hr\cdot\bfk\ \ \sigma_2\cdot\bfk'\times\hr+
B_2C_1\ \bfk'\cdot\hr\ \sigma_1\cdot\hr\times\bfk)\ g^-(r)
-(B_1C_2\ \sigma_2\cdot\bfk'\times\bfk+B_2C_1\ \sigma_1
\cdot\bfk'\times\bfk)\ \frac{g'(r)}{r} $$
$$ -C_1C_2\ \left[(\sigma_1\cdot\hr\times\bfk\ 
\sigma_2\cdot\bfk'\times\hr) g^-(r)+(\sigma_1\cdot\bfk'\ 
\sigma_2\cdot\bfk-\sigma_1\cdot\sigma_2\ \bfk\cdot\bfk')
\ \frac{g'(r)}{r}\right] $$ 
where $g^-(r)\equiv g''(r)-g'/r$.  

Terms proportional to $g'(r)/r$ are ``quantum'' in origin and fall off
as $1/r^2$ for large distances.  For large values of $r$ also

$$g(r)\rightarrow \frac{e^{i\kappa r}}{r};\ \ 
g''(r)\rightarrow -\kappa^2\frac{e^{i\kappa r}}{r}; \ \ 
g'(r)\rightarrow i\kappa\frac{e^{i\kappa r}}{r}.$$

The spin-independent terms will be diagonal in the initial and final
states but we must take the expectation values of the the spin
operators for the other terms.  We will use the singlet-triplet
representation for the present problem since the initial state is a
pure triplet. The matrix elements needed for the spin amplitudes are
given in the appendix.

\subsection{Isospin of the deuteron}

In order to include the effect the definite isospin of the deuteron we 
can write the amplitude as

\eq
M=<pp|\sum_{i\ne j,j=1,2}
e^{i(\bfk_{\pi}\cdot\bfr_j-\bfk'_{\pi}\cdot\bfr_i)}
\int d\bfq \frac{f_i(\bfq,\bfk')f_j(\bfk,\bfq)
e^{i\bfq\cdot(\bfr_i-\bfr_j)}}{\bfq^2-k^2}|D>
\qe
where the operators $f_i$ are the pion-nucleon amplitudes in
spin and isospin space and the bras and kets refer to isospin states
only. We have included the initial and final spatial states of the
pion but not the final state of the two protons. Since 
there are two orders of the scattering
possible and there are two terms in the isospin expansion of the
deuteron wave function there are 4 terms in this expression, each
of the type presented in the previous section. 

Thus, since an operator cannot act on the same particle successively
the effect of one term on the isospin part of the  deuteron wave function 
is
\eq
<p_1|<p_2|(f_2f_1+f_1f_2)\frac{1}{\sqrt{2}}(|p_1>|n_2>-|n_1>|p_2>)
\qe
where each product of $f$'s can be decomposed into terms from Eq.  
\ref{expan} consisting of a constant multiplying an operator.  We make the
simplifying approximation that the amplitudes for the elastic scatterings
and charge exchanges depend only on the pion momentum (neglecting the
nucleon motion).  Thus, we assume all nucleons at rest for the purpose of
the evaluation of the $\pi$N amplitudes only.

These considerations allow us to write this in the form
\eq
f_2^x(k'_{\pi})f_1^+(k_{\pi})-f_2^0(k'_{\pi})f_1^x(k_{\pi})
+f_1^0(k'_{\pi})f_2^x(k_{\pi})-f_1^x(k'_{\pi})f_2^+(k_{\pi}).
\qe
Considering, term by term, the components in Eq. \ref{expan} which
have the form of constants times operators and taking the expression
for double scattering of a generic operator to be represented by
$R(\bfr,k)$ and a generic constant amplitude for the corresponding
term to be $D$ we can write

\eq
\left[D^x(k'_{\pi})D^+(k_{\pi})-D^0(k'_{\pi})D^x(k_{\pi})\right]
R(\bfr,q_1)e^{i(\bfk_{\pi}+\bfk'_{\pi})\cdot \bfr /2}
\qe
$$
+\left[D^0(k'_{\pi})D^x(k_{\pi})-D^x(k'_{\pi})D^+(k_{\pi})\right]
R(-\bfr,q_2)
e^{-i(\bfk_{\pi}+\bfk'_{\pi})\cdot \bfr /2}
$$

\eq = \left[D^x(k'_{\pi})D^+(k_{\pi})-D^0(k'_{\pi})D^x(k_{\pi})\right]
\left[R(\bfr,q_1)e^{i(\bfk_{\pi}+\bfk'_{\pi})\cdot \bfr /2}
-R(-\bfr,q_2)e^{-i(\bfk_{\pi}+\bfk'_{\pi})\cdot \bfr /2}\right]
\label{isoact}\qe
where $\bfq_i=\bfk_{\pi}-\bfk_i$ is the intermediate momentum of the
propagating pion in each case.

The subtraction of the strength of the two possible interactions
represented by the differences of the multiplying constants $D$ does
not depend on the spatial coordinates. The minus sign can be traced to
the isospin character of the deuteron.

\subsection{Phases from the proton-proton final state}

For the spatial final state wave function of the two protons we have

\eq
e^{-i(\bfk_1\cdot\bfr_1+\bfk_2\cdot\bfr_2)}\pm
e^{-i(\bfk_1\cdot\bfr_2+\bfk_2\cdot\bfr_1)}\rightarrow
e^{i(\bfk_2-\bfk_1)\cdot\bfr/2}\pm
e^{-i(\bfk_2-\bfk_1)\cdot\bfr/2}
\qe
where the plus sign corresponds to a singlet final state and the
minus sign to the triplet final state.

Taking into account the conservation of momentum $\bfk_{\pi}
=\bfk'_{\pi}+\bfk_1+\bfk_2$ we can combine the phase factor of
the two terms in Eq \ref{isoact} (dropping  the overall multiplying
constant for the moment) to find

\eq
\left[e^{i\bfr\cdot\bfq_1}\pm e^{i\bfr\cdot\bfq_2}\right]R(\bfr,q_1)
-\left[e^{-i\bfr\cdot\bfq_1}\pm 
e^{-i\bfr\cdot\bfq_2}\right]R(-\bfr,q_2).
\qe

The operators in Eq. \ref{expan} have a definite character in 
the parity of $\bfr$, either even or odd. The next step in computing the 
quantum matrix element would be to integrate over this vector thus picking 
out the matching terms in the final state nucleon wave function. Since we 
are ``correcting'' for the quantum effect event by event (and each event 
has a definite value of $\bfr$) we cannot proceed to this integration step 
but we will keep only those terms which would survive this integration.
This leads us to the following array which must be applied term by
term.

{\small
$$
\begin{array}{|l|c|c|}
\hline
{\rm Spin}&{\rm Triplet\ to\ Triplet}&{\rm Triplet\ to\ Singlet}\\
\hline
{\rm Even}&
[\cos(\bfr\cdot\bfq_1)-\cos(\bfr\cdot\bfq_2)][R(\bfr,q_1)+R(\bfr,q_2)]
&[\cos(\bfr\cdot\bfq_1)+\cos(\bfr\cdot\bfq_2)][R(\bfr,q_1)-R(\bfr,q_2)]
\\
\hline
{\rm Odd}&
i[\sin(\bfr\cdot\bfq_1)-\sin(\bfr\cdot\bfq_2)][R(\bfr,q_1)+R(\bfr,q_2)]
&i[\sin(\bfr\cdot\bfq_1)+\sin(\bfr\cdot\bfq_2)][R(\bfr,q_1)-R(\bfr,q_2)]\\
\hline
\end{array}
$$
}

\section{Results and conclusions}

The quantum effects on double scattering discussed were implemented in the
calculation by computing a weight corresponding to each event.
Calculations were performed with $4\times 10^8$ cascades.

The poor agreement of the obtained potential with a semi-realistic
nucleon-nucleon may worry some, and with good reason.  However, it seems
to be necessary in order to obtain some even more important conditions
in a classical simulation. First, the density distribution of nucleons
must be correct or else the magnitude of the cross section and estimates
of multiple scattering will be wrong. Even the early INC codes did this
(more or less) correctly. Second, Fermi momentum must be included. Without
this physical effect the coincidence spectra would appear as a series
of spikes. The correct degree of smearing is very important. Third, energy
must be conserved and definite. If one simply includes the motion of the 
nucleons without adding a potential to compensate the kinetic energy of
the nucleons that corresponds to the Fermi motion, the deuterium nucleus
will not have a definite energy and such features as the Jacobian peaks
would be washed out. Thus, these three conditions are absolutely
essential for the present calculation.  The selection of any two
implies the third, there is no choice: we are left with a specified
potential.  

The non-realistic nature of this potential mainly affects the single
scattering through distortion of the distribution of the final-state
momentum of the spectator particle.  Since single scattering is
largely eliminated by the momentum thresholds we do not expect a
large problem. In those cases in which there remains a significant
contribution from single scattering, errors may occur.  We believe
that we have taken the correct compromise for this particular set of
observables. For another case (one in which very low energy protons
were detected, for example) it might be more appropriate to choose a
realistic potential at the expense of the Fermi momentum distribution
or the correct density.

\begin{center}
\begin{figure}[p]
\epsfysize=185mm
\epsffile{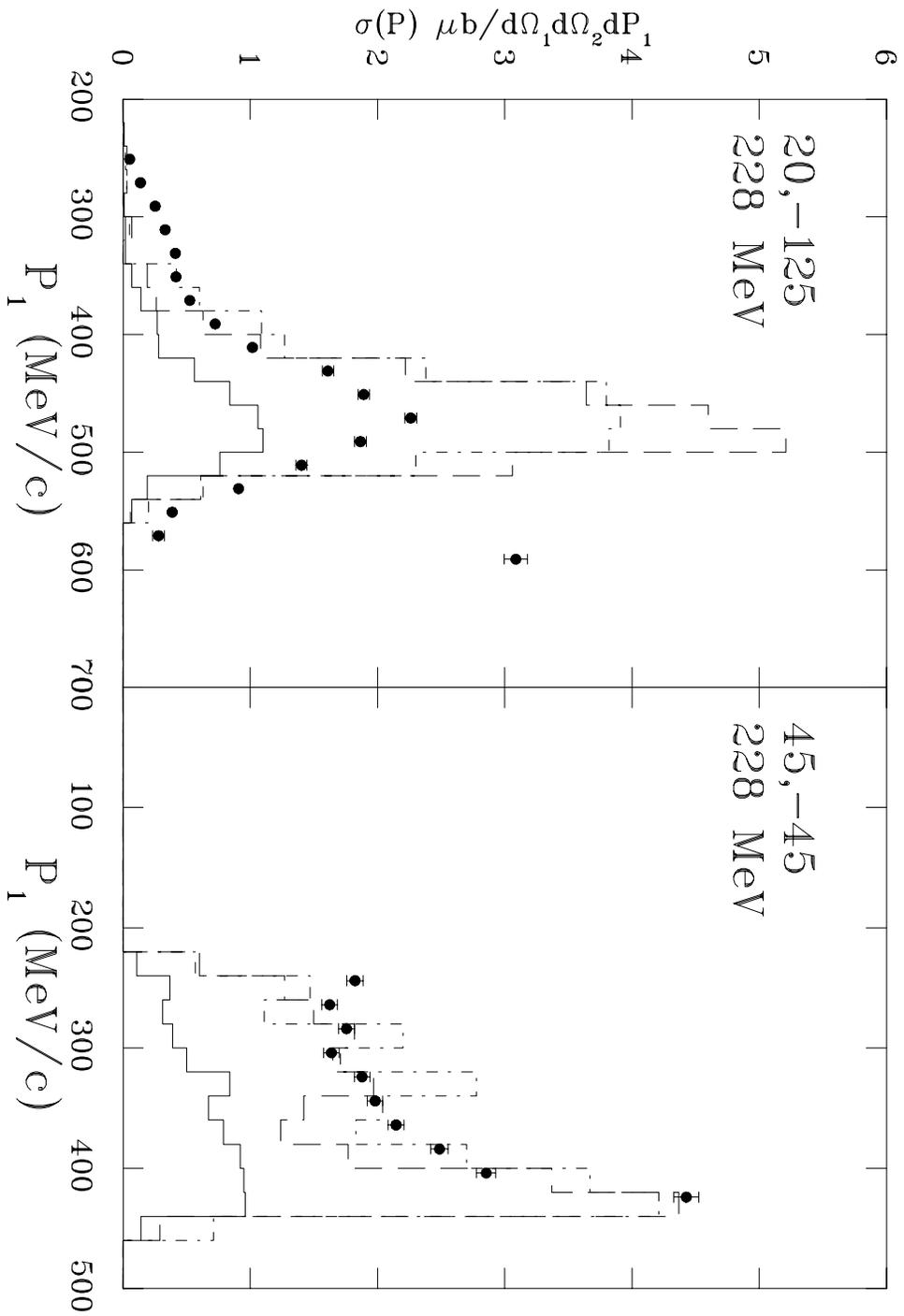}
\caption{Double scattering comparing classical (dashed line), fully 
quantum (solid line) and partial quantum effects without the coherent 
deuteron wave function (dash-dot line) at 228 MeV.}
\label{cohqnqa}
\end{figure}
\end{center}

Figure \ref{cohqnqa} illustrates for a typical pair of angles that the
isospin correction is the most important one.  The phase correction is
much smaller. The constants, $D$, in Eq. \ref{isoact} tend to cancel. If
the amplitude were completely dominated by the 33 resonance, there would
be a constant reduction factor. That dominance is not so pronounced at
these energies but there is still a significant cancellation in many
cases.

\begin{figure}[p]
\begin{center}
\epsfysize=185mm
\epsffile{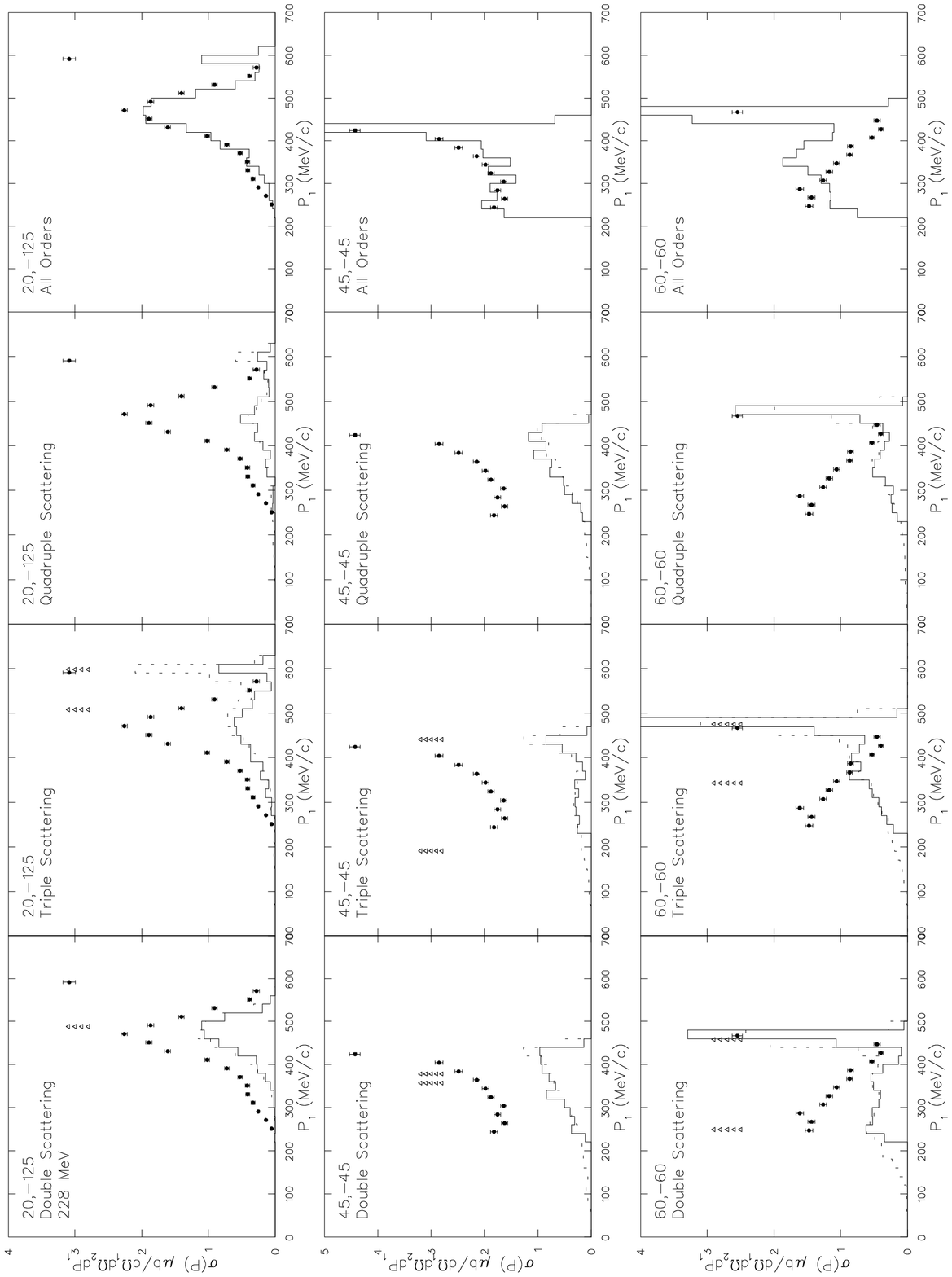}
\caption{Comparison of orders of multiple scattering and the total
at 228 MeV. The single scattering for (45\de,--45\de) is shown in
Fig. \protect{\ref{sbernfca6}}.}
\label{bernms228}
\end{center}
\end{figure}

\begin{figure}[p]
\begin{center}
\epsfysize=185mm
\epsffile{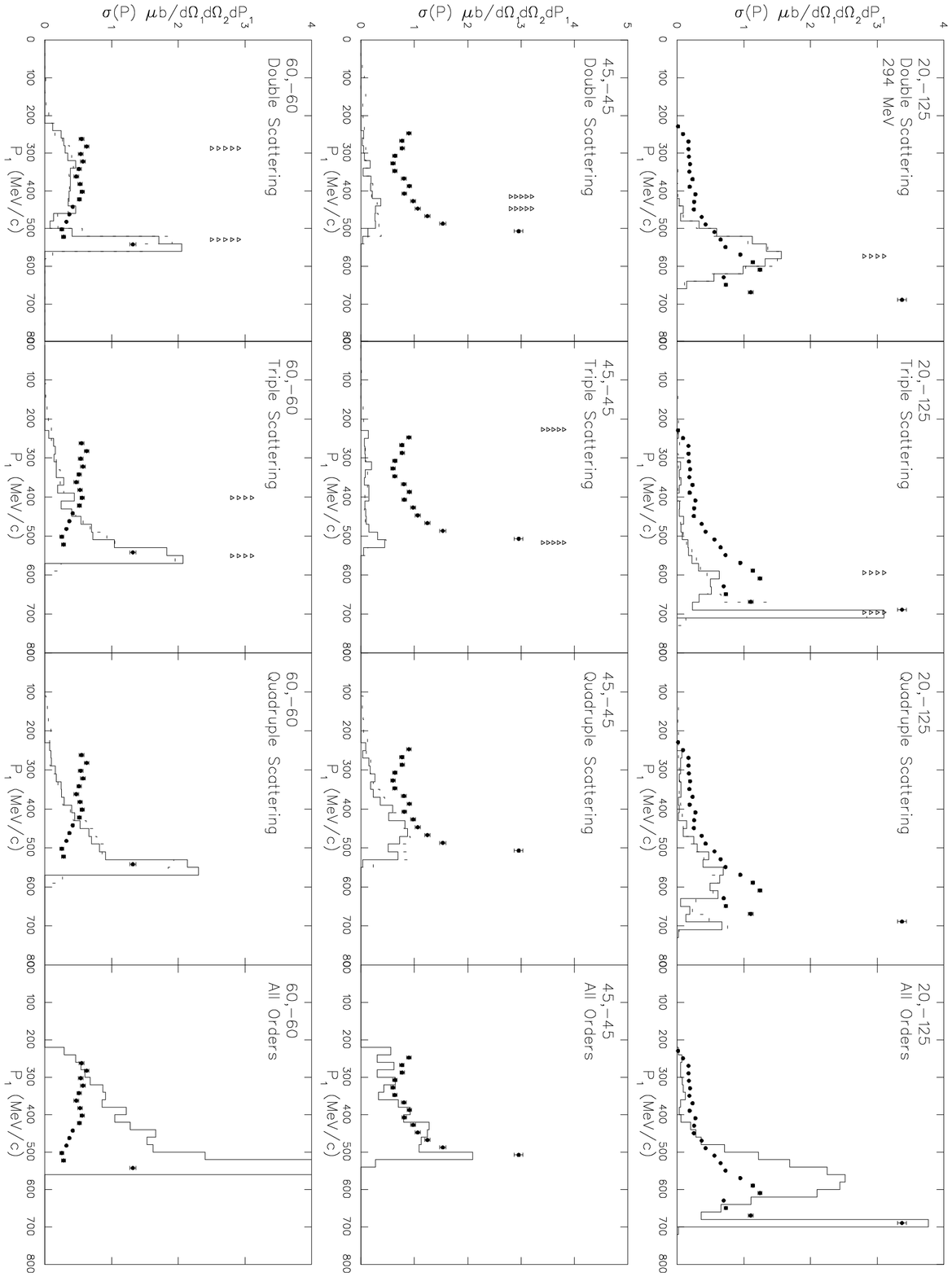}
\caption{Comparison of orders of multiple scattering and the total
at 294 MeV. The single scattering for (45\de,--45\de) is shown in
Fig. \protect{\ref{sbernfca6}}.}
\label{bernms294}
\end{center}
\end{figure}

Figures \ref{bernms228} and \ref{bernms294} show the various orders of
multiple scattering beyond single.  The interactions in the cascade were
stopped at fourth order so quadruple scattering really includes all 
higher orders which would have occurred if allowed to continue. We have
seen that the multiple scattering goes to higher orders at 294 MeV than
at 228 MeV. One possible reason for this is that the absorption is less
at the higher energy. When the energy is degraded by collisions the
absorption becomes larger and truncates the multiple scattering. This
may be one reason why we have so much multiple scattering.

\begin{center}
\begin{figure}[p]
\epsfysize=185mm
\epsffile{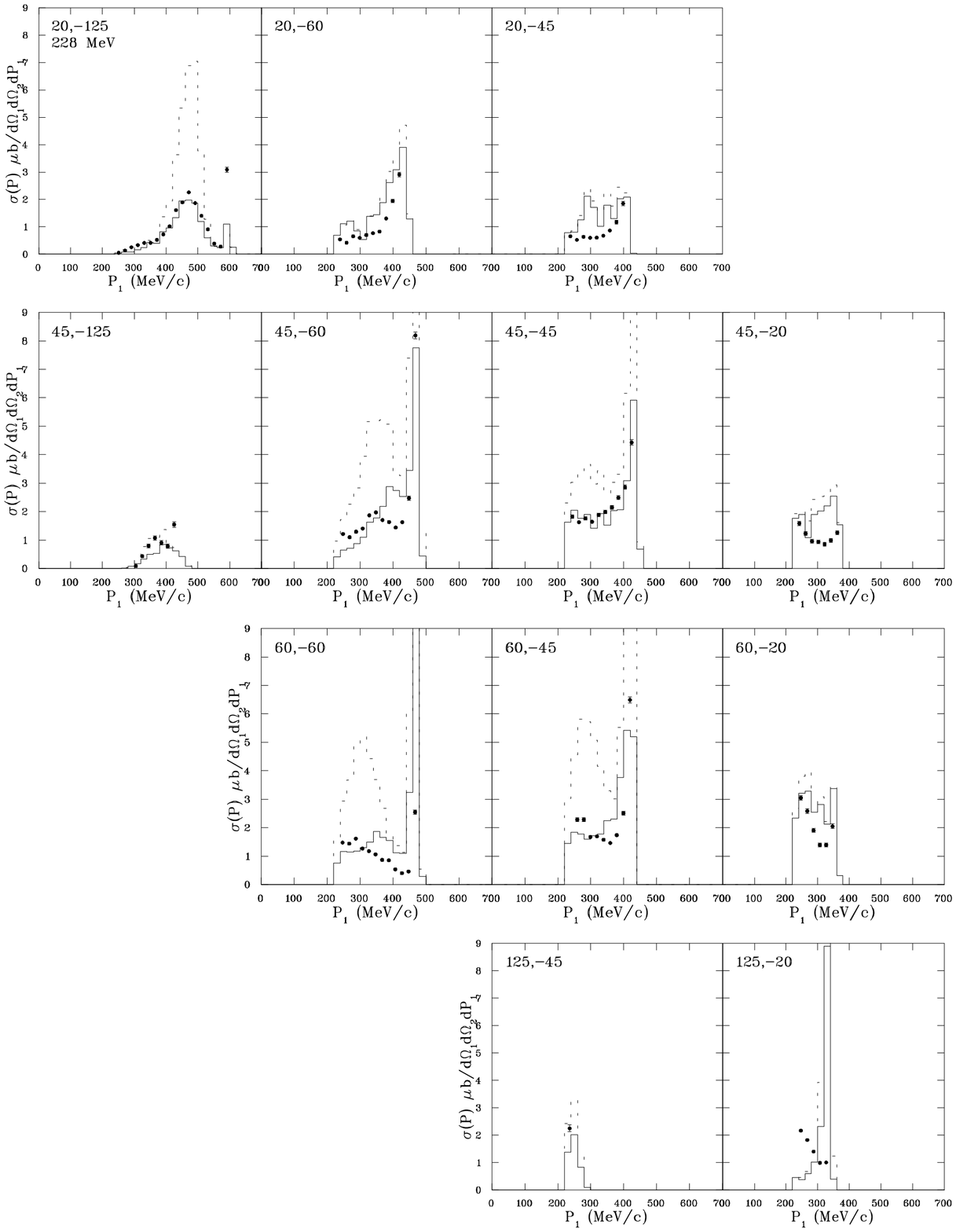}
\caption{Total scattering at 228 MeV with (solid) and without (dashed)
quantum corrections.}
\label{tot228}
\end{figure}
\end{center}

\begin{center}
\begin{figure}[p]
\epsfysize=185mm
\epsffile{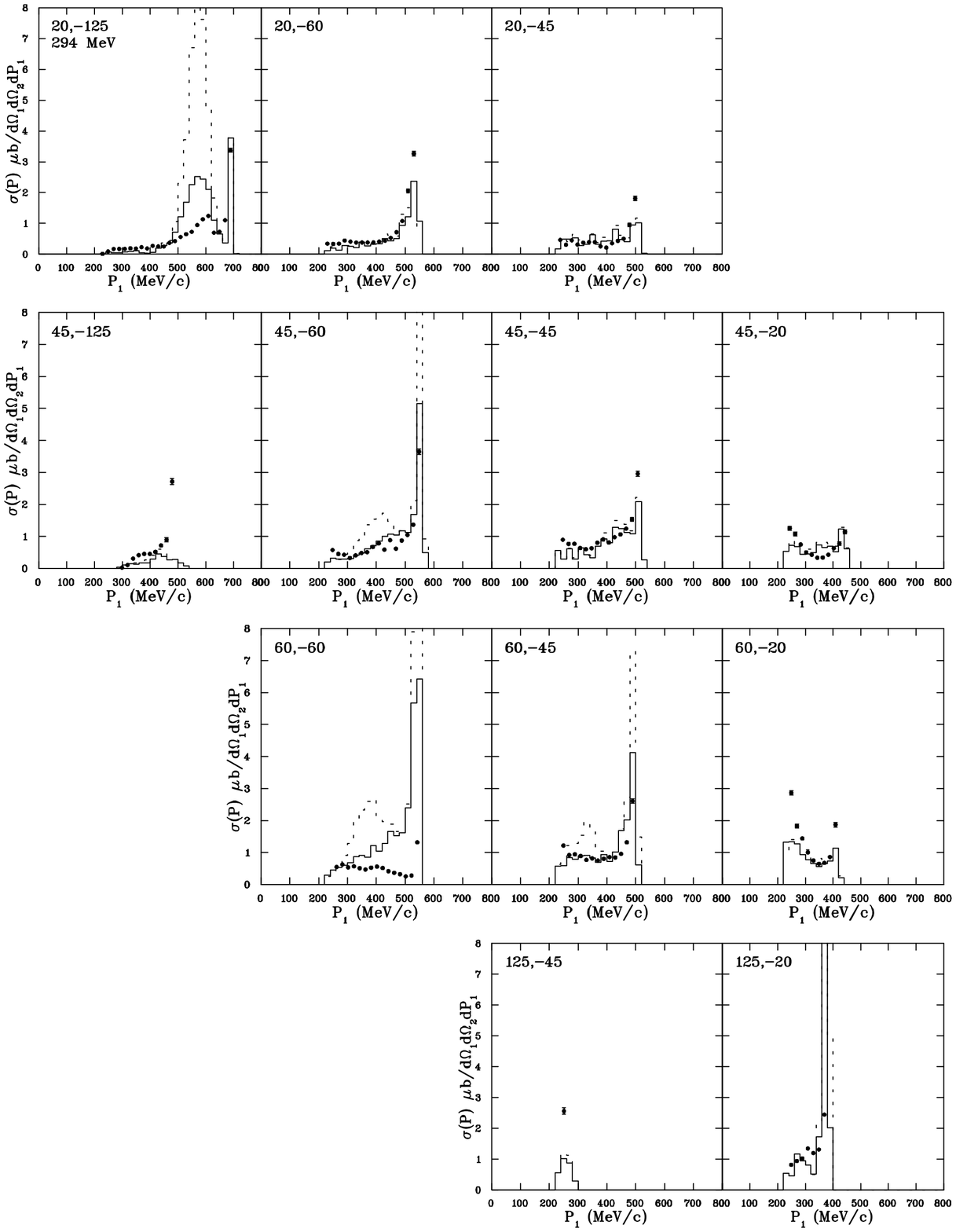}
\caption{Total scattering at 294 MeV with (solid) and without (dashed)
quantum corrections.}
\label{tot294}
\end{figure}
\end{center}

Figures \ref{tot228} and \ref{tot294} give the results for all angular
pairs with and without quantum corrections in the double scattering.
We have seen that the quantum effects included (especially the isospin
one) give a large decrease in the double scattering cross section
which carries over into the total as well.  We see that the agreement
with the data at 228 MeV is generally good with the possible exception
of the counter pairs (60\de,--45\de) and (60\de,--60\de) where the
cross section is overestimated in the mid-momentum range.  At 294 MeV
the agreement is excellent except for a substantial overestimate for
the pairs (60\de,--60\de) and (20\de,--125\de).  Much, but not all, of
the overestimate (in the first counter pair at least) can be
attributed to third and higher order scatterings which suggests that
in some cases the model overestimates these contributions.  One
possible reason could be that the quantum corrections have not been
made to these orders. We anticipate that such corrections would go
again in the direction of reducing the cross section but their 
inclusion, though possible, is beyond the scope of the present work.

The comparison for the counter pair (125\de,--20\de) at 228 MeV is 
puzzling. The data reverses its trend from the same pair at 294 MeV
while the calculation gives the same general form.

We thank R. Tacik for supplying us with tables of the data
in electronic form.  JPD wishes to thank the Dept. of Physics of New 
Mexico State University for its hospitality and partial support and 
WRG expresses appreciation for the same to the Universit\'e Paris 7-
Denis Diderot. This work was supported by the National Science 
Foundation under contract PHY-0099729.

\newpage

\begin{center}{\bf Appendix}\end{center}

\begin{appendix}

\section{Spin matrix elements}

With the definitions
\eq \bfc=\bfa\times\bfb \qe
\eq \az=a^z;\ \ \apo=\frac{1}{\sqrt{2}}(a^x+ia^y);\ \ 
 \amo=-\frac{1}{\sqrt{2}}(a^x-ia^y), \qe
the matrix elements of
$<S'S_z'|\sigma_1\cdot\bfa\ \ \sigma_2\cdot\bfb |SS_z>$ can
be written
$$ \begin{array}{cccccccc}
&&\vline&(0,\ \ 0)&(1,\ -1)&(1,\ \ 0)&(1,\ +1)&(S',S_z')\\ \hline 
(S,S_z)&(0,\ \ \ 0)&\vline&-\bfa\cdot\bfb&-i\cmo&i\cz&-i\cpo \\
&(1,\ -1)&\vline&-i\cpo&\az\bz&-\az\bpo-\apo\bz&2\apo\bpo\\
&(1,\ \ \ 0)&\vline&-i\cz&\az\bmo+\amo\bz&\bfa\cdot\bfb-2\az\bz&
\apo\bz+\az\bpo \\
&(1,\  +1)&\vline&-i\cmo&2\amo\bmo&-\amo\bz-\az\bmo&\az\bz \\
\end{array} $$

We need this matrix twice, once for $\bfa =\bfk$; $\bfb =\bfk '$ 
and
once for $\bfa =\hr \times\bfk$; $\bfb =\bfk '\times\hr$.  For the
second case we can write $\bfc =-(\hr\cdot\bfk\times\bfk ')\hr$.

For a single spin operator we have
$$ \begin{array}{cccccc}
&\vline&(0,\ \ 0)&(1,\ -1)&(1,\ \ 0)&(1,\ +1)\\ \hline 
(0,\ \ \ 0)&\vline&0&-\amo&\az&-\apo\\
(1,\ -1)&\vline&\apo&-\az&\apo&0\\
(1,\ \ \ 0)&\vline&\az&-\amo&0&\apo\\
(1,\ +1)&\vline&\amo&0&-\amo&\az\\
\end{array} $$
for the matrix elements of $<S'S_z'|\sigma_1\cdot\bfa|SS_z>$ and

$$ \begin{array}{cccccc}
&\vline&(0,\ \ 0)&(1,\ -1)&(1,\ \ 0)&(1,\ +1)\\ \hline 
(0,\ \ \ 0)&\vline&0&+\amo&-\az&\apo\\
(1,\ -1)&\vline&-\apo&-\az&\apo&0\\
(1,\ \ \ 0)&\vline&-\az&-\amo&0&\apo\\
(1,\ +1)&\vline&-\amo&0&-\amo&\az\\
\end{array} $$
for the matrix elements of $<S'S_z'|\sigma_2\cdot\bfa|SS_z>$.

\end{appendix}

\end{document}